\documentclass[conference]{IEEEtran}
\IEEEoverridecommandlockouts
\usepackage{cite}
\usepackage{amsmath,amssymb,amsfonts}
\usepackage{algorithmic}
\usepackage{graphicx}
\usepackage{romannum}
\usepackage{textcomp}
\usepackage{xcolor}
\usepackage{mybeamer}
\def\BibTeX{{\rm B\kern-.05em{\sc i\kern-.025em b}\kern-.08em
    T\kern-.1667em\lower.7ex\hbox{E}\kern-.125emX}}
\newtheorem{definition}{Definition}

\renewcommand{\bl}[1]{\textcolor{blue}{#1}}

\makeatletter
\newcommand{\removelatexerror}{\let\@latex@error\@gobble}
\makeatother

\usepackage{xcolor}

\begin{document}

\title{Mining Message Flows using Recurrent Neural Networks for System-on-Chip Designs}

\author{\IEEEauthorblockN{Yuting Cao}
\IEEEauthorblockA{
\textit{Synopsys}\\
Mountain View, CA, USA \\
yutingc@synopsys.com}
\and
\IEEEauthorblockN{Parijat Mukherjee, Mahesh Ketkar, Jin Yang}
\IEEEauthorblockA{
\textit{Intel}\\
\{Hillsboro, OR, Folsom, CA, Hillsboro, OR, \}USA \\
\{parijat.mukherjee, mahesh.c.ketkar, jin.yang\}@intel.com}
\and
\IEEEauthorblockN{Hao Zheng}
\IEEEauthorblockA{
\textit{CSE, U of S Florida}\\
Tampa, FL, USA \\
haozheng@usf.edu}
}

\def\checkmark{\tikz\fill[scale=0.4](0,.35) -- (.25,0) -- (1,.7) -- (.25,.15) -- cycle;} 
\maketitle

\begin{abstract}
Comprehensive specifications are essential for various activities across the entire validation continuum for system-on-chip (SoC) designs. However, specifications are often ambiguous, incomplete, or even contain inconsistencies or errors.
This paper addresses this problem by developing a specification mining approach that automatically extracts sequential patterns from SoC transaction-level traces such that the mined patterns collectively characterize system-level specifications for SoC designs.  This approach exploits long short-term memory (LSTM) networks trained with the collected SoC execution traces to capture sequential dependencies among various communication events.  Then, a novel algorithm is developed to efficiently extract sequential patterns on system-level communications from the trained LSTM models.  Several trace processing techniques are also proposed to enhance the mining performance.  
We evaluate the proposed approach on simulation traces of a non-trivial multi-core SoC prototype.  Initial results show that the proposed approach is capable of extracting various patterns on system-level specifications from the highly concurrent SoC execution traces. 

%



\end{abstract}

\section{Introduction}
\label{sec:intro}
Well-formed and comprehensive specifications are essential for various system-on-chip (SoC) validation activities.  In practice, such specifications are often ambiguous,  incomplete,  or  even  contain  inconsistencies  or  errors.
Moreover, as a SoC design gradually progresses across design stages, the connection between the original specifications and the design implementation may become imprecise and disjoint.
The lack of accurate specifications may lead to potential misunderstandings of the design behavior, and cause unintentional misbehavior to be implemented.  It also hinders effective debug process.
In order to address those challenges, an automatic system specification extraction method is crucial for effective SoC validation.

In the literature, many existing methods allow extracting specifications from system execution traces. However, they are mainly designed for software, and cannot be directly applied to SoC designs. In this paper, we consider the concurrent nature of SoC designs, where execution traces are results from executing a number of message flows in parallel.   
In the proposed specification mining approach, the state-of-art LSTM neural networks,  which are effective at capturing sequential dependencies, are trained with the SoC execution traces.  Subsequently, sequential patterns are automatically extracted from the trained LSTM models, where message flow specifications can be formed from the mined sequential patterns.

The contribution of this paper is a novel sequential pattern mining framework that automatically extracts sequential patterns from inherently concurrent SoC execution traces.  These patterns may be recurrent in individual traces, and repeat themselves in multiple different traces.  Even though the type of execution traces and the type of mined sequential patterns are considered in separation in some previous work, to the best of our knowledge, this is the first work that both features are considered in a single mining framework.

This paper is organized as follows.  Section~\ref{sec:rel-work} reviews the existing related work. Section~\ref{sec:background} and \ref{sec:seq_pat} introduce the basic definitions and concepts. Section~\ref{sec:method} presents our specification mining framework. Section~\ref{section:results} discusses the experimental results, and Section~\ref{sec:conclude} concludes this paper.
\section{Related Work}
\label{sec:rel-work}

Whereas numerous researches are conducted to extract sequential patterns from traces of software programs \cite{beschastnikh2011leveraging,biermann1972synthesis,lorenzoli2008automatic,lou2010mining,liu2013automatic, chang2010automatic}, they are not specifically designed for concurrent programs. 
The work in \cite{chang2010automatic} extracts assertions automatically by applying data mining methodologies to simulations traces of digital hardware designs. 
The work in \cite{liu2013automatic} uses the similar concept on transaction-level traces. 
However, both approaches cannot handle concurrency that commonly exists in complex SoC execution traces.
The approaches proposed in~\cite{beschastnikh2014inferring,lou2010mining,yu2016cloudseer,du2017deeplog, le2018deep} target traces of the concurrent system.
They show that common model-inference algorithms are not directly applicable to distributed concurrent system executions. 
CloudSeer, proposed in~\cite{yu2016cloudseer}, is a specification inference approach that can be applied to anomaly detection.
It requires a set of log files with repeated executions of only one single task, which makes it infeasible for SoC traces. For the same reason, works in \cite{beschastnikh2014inferring,lou2010mining} are also limited. 

In practice, a highly concurrent program often has multiple tasks running concurrently, with their log entries interleaved with each other.
A recent work, BaySpec~\cite{mrowca2019learning}, uses a dynamic mining approach to extracts formal specifications from Bayesian models. 
This work, unfortunately, cannot be applied to SoC traces as it requires clean trace slicing (requires functional segmentation), which is impractical.
The work in \cite{du2017deeplog} suffers the similar issue. 
Another work in \cite{le2018deep} presents an approach that constructs a Prefix Tree Acceptor (PTA) from software traces. It then extracts and combines patterns from the constructed PTA.  It also requires traces be properly sliced in order to construct the PTA.
\section{PRELIMINARIES}
\label{sec:background}



\begin{figure}[t]
\begin{center}
\includegraphics[width=3in]{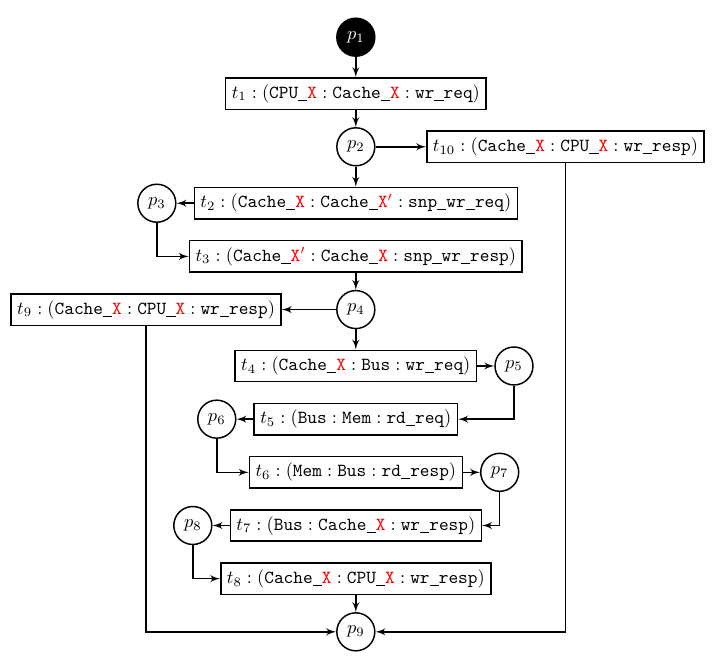}
\caption{LPN formalization of a CPU write protocol.  }
\label{fig:ex}
\end{center}
\end{figure}

In architectural documents, system-level protocols are often represented as message flows, therefore they are also referred to as system flows in this paper. 
\begin{definition} A \textbf{system flow} is defined as a tuple {${F} = (P, T, E, L)$} where {$P$} is a finite set of \emph{places}, $T$ is a finite set of \emph{transitions}, $E$ is a finite set of \emph{events}, and $L: T \rightarrow E$ is a labeling function that maps each transition $t \in T$ to an event $e \in E$.
\end{definition}
An SoC typically implements several flows denoted as $\vec{F}$.  In this paper, $F_i \in \vec{F}$ denotes one such flow.  An example of memory write flow from CPUs is shown in Fig.~\ref{fig:ex} where each transition is labeled with an index $t_x$ and an event of the form {\tt (A:B:C)}

For each transition $t \in T$, its preset, denoted as $\bullet{t} \subseteq P$, is the set of places connected to $t$, and its postset, denoted as $t\bullet \subseteq P$, is the set of places that $t$ is connected to.  A state $s \subseteq P$ of a flow is a subset of places marked with tokens.  Each flow has two special states: 
$s_0 \subseteq P$ is the set of initially
marked states, also referred to as the \emph{initial state}, and the end state $s_{\perp} \subseteq P$ is the set of end states not going to any transitions.  Each flow is associated with one \emph{start} and several \emph{end} events.  An event $e \in E$ is a start event if $e = L(t)$ and $\bullet t \subseteq s_0$.  An event $e \in E$ is an end event if $e = L(t)$ and $t \bullet \subseteq s_{\perp}$. In Figure~\ref{fig:ex}, $s_0 = \{p_1\}$, and $s_{\perp} = \{p_9\}$, its start event is 
$t_1$, and its end event is the one labeled for transitions $t_8, t_9$ and $t_{10}$.  The occurrence of a start event indicates the beginning of a flow execution, while the occurrence of an end event indicates the complete of a flow execution. 
A transition $t$ can be fired in a state $s$ if $\bullet t \subseteq s$.  Firing $t$ causes the labeled event to be emitted, and leads to a new state $s^\prime = (s - \bullet t) \cup t \bullet$. Therefore, executing a flow induces a sequence of events.  Execution of a flow completes if its $s_{\perp}$ is reached.  



\begin{definition}
An {\bf execution} of a flow $F$ is a sequence of events $(e_0, e_1, \ldots, e_n)$ such that there is a sequence of transition firings $(t_0, t_1, \ldots, t_n)$ of $F$ where the following conditions hold.
\begin{itemize}
\item $\forall 0\leq i \leq n,\ e_i = L(t_i)$,
\item $\forall 0 < i \leq n,\ \bullet t_i \subseteq t_{i-1}\bullet$,
\item $s_0 \subseteq \bullet t_0$, and $t_n \bullet \subseteq s_\perp$.
\end{itemize}
\end{definition}

The above definition indicates that an execution of $F$ is the results of a sequence of transition firings from the initial state to the end state where transition $t_{i}$ is causally dependent on $t_{i-1}$.
For example, the flow in Figure~\ref{fig:ex} has three executions: $\{t_1, t_{10}\}$, $\{t_1, t_2, t_3,t_9\}$ and $\{t_1, t_2, t_3, t_4, t_5, t_6, t_7, t_8\}$.  Note that the transition numbers are used to represent events in the above example for simplicity.


\begin{definition} 
During an execution of an SoC design, instances of flows $\vec{F}$ that it implements are executed.  Suppose that the set of flow instances executed is $\{F_{i,j}~|~F_i \in \vec{F}\}$.
An SoC execution yields a \textbf{trace $\rho$}  
\[
\rho = (\mathcal{E}_0, \mathcal{E}_1, \ldots, \mathcal{E}_n)
\]
where $\mathcal{E}_i = \{e_{i,0},\ ...e_{i,k}\}$ is a set of events executed at time $i$, and $e_{i,\ast} \in \bigcup E_{i,j}$ for every $e_{i,\ast} \in \mathcal{E}_i$.
\end{definition}

Intuitively, an SoC execution trace is the result of parallel execution of multiple instances of different flows.  It is a sequence of sets of events where orderings of events in the same set of a trace are indistinguishable. 

Given a trace $\rho$ and two events $e_i$ and $e_j$, we define $e_i < e_j$ if $e_i \in \mathcal{E}_i$, $e_j \in \mathcal{E}_j$, and $i < j$.  This notation is naturally extended to sequences of events. 

In a system flow, an event is a tuple $({\tt src: dest: cmd}$ $ {\tt\{data\_field\}})$ where ${\tt cmd}$ is a command sent from a source component ${\tt src}$ to a destination component ${\tt dest}$, and  ${\tt data\_field}$ represents the payload information carried by the event.  It can be ${\tt addr}$ that carries the memory address at the target block where ${\tt cmd}$ is applied, or other information like ${\tt sequence\_ID}$ and ${\tt tag}$ to identify an event.  The information defining an event can be classified as {\em static} including ${\tt src}$, ${\tt dest}$ and ${\tt cmd}$,  and {\em dynamic} including various ${\tt data\_field}$ which is generated  during SoC execution.  

 In this paper, we consider all the static information of an event for mining.  Without loss of generality, we also consider dynamic information ${\tt addr}$ of ${\tt data\_field}$ for mining as well.  All other types of dynamically generated information are not considered.  However, users can selectively use them in the mining process to achieve better results.

\section{Sequential Patterns}
\label{sec:seq_pat}



This paper considers mining message flows implemented in SoC on-chip fabric.  Since these flows are implemented in hardware, mined flows, if valid, must be invariant on different execution traces.  In particular, this paper considers mining sequential patterns, each of which corresponds to an execution of a message flow.  


\label{sec:rules}
\begin{definition}
A {\bf sequential pattern $p$} is a sequence of events such that \begin{itemize}
    \item It consists of at least two events, and 
    \item All its events are \emph{unique}.
\end{itemize}
\end{definition}


If ground truth patterns are known, validity of mined patterns can be defined with respect to the ground truth patterns.
\begin{definition}
A mined pattern $p_m$ is {\bf valid} if there is a ground truth (GT) pattern $p_t$ such that for every pair of events $e_i$ and $e_j$ in $p_m$, $e_i < e_j$ in $p_m$ if  $e_i < e_j$ in $p_t$.
A mined pattern $p_m$ is {\bf invalid} if it is not valid.
\end{definition}

In the simple example below, $p_m$ is a mined pattern.  $p_m$ is valid with respect to $p_t$ as sequential dependencies between any pair of events in $p_m$ also exists in $p_t$.  
\[
\begin{array}{rl}
p_m: &	(\bl{0}, \bl{13}, \bl{15}, \bl{23})\\
p_t:	& ({\color{blue}0}, 8, 12, {\color{blue}13}, {\color{blue}15}, {\color{blue}23}, 24, 25)\\
\end{array}
\]



Next, we show an interesting property that helps identify valid patterns. 
It captures the cause-effect relation between two events, and it is based on the following observation: \emph{any event in an execution trace is generated by a component in a SoC design in reaction to an input event}. 
\begin{definition} 
\label{def:causality}
For two events $e_i$ and $e_j$ in a trace such that  $e_i < e_j$,  they satisfy the {\bf causality} property if 
\[
e_i.{\tt dest} = e_{j}.{\tt src}. 
\]
\end{definition}

An execution in a message flow specifies a sequence of such relations.  Therefore, for a pattern $(e_0, e_1, \ldots, e_k)$ where $0 \leq i < k$ to be valid, every two consecutive events $e_i$ and $e_{i+1}$ in the pattern must satisfy the causality property.


\section{Mining Framework}
\label{sec:method}

\begin{figure*}[tb]
    \centering
    \includegraphics[width=.7\textwidth]{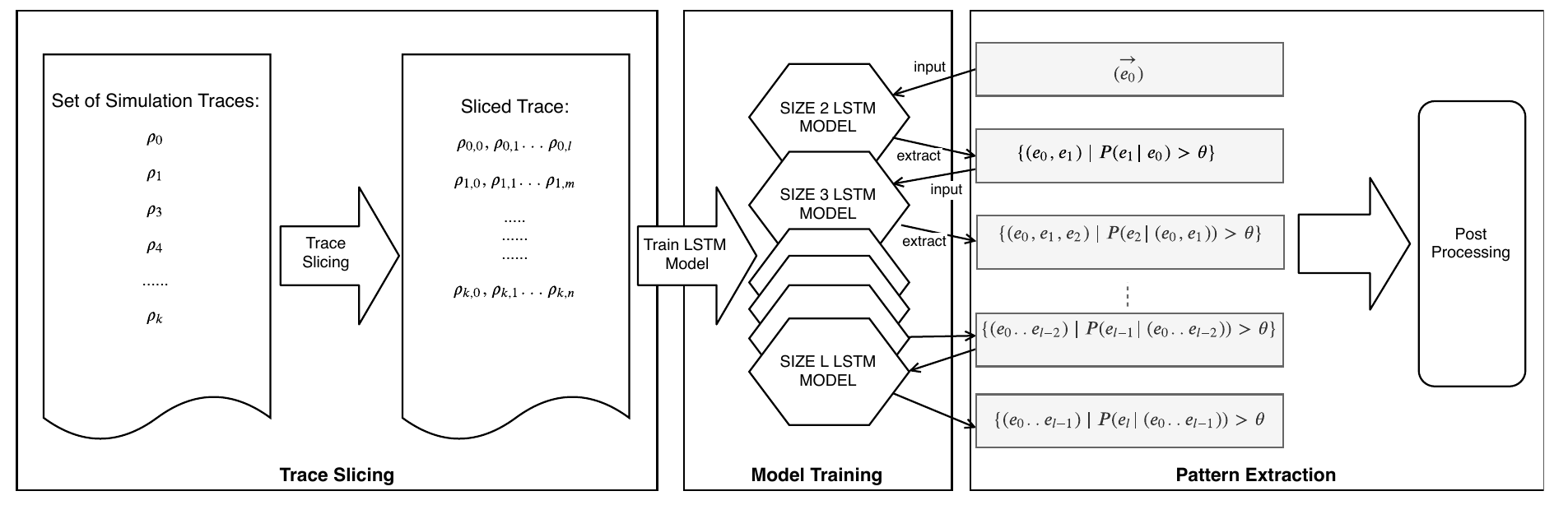}
    \caption{Overview of the sequential pattern extraction approach architecture}
    \label{fig-overview}
\end{figure*}

Figure~\ref{fig-overview} shows the overview of the proposed mining framework.  It accepts a set of SoC execution traces, and mines sequential patterns from those traces in three consecutive stages as described below.




\subsection{Trace Processing} 
\label{sec-slicing}
Modern SoC model are often highly concurrent, and can execute multiple flows simultaneously. As a result,
it is possible that two unrelated events may follow a temporal dependency in execution traces.  Since our mining framework infers causality dependencies from temporal ones, those false temporal dependencies can lead to many invalid patterns to be mined. 
To address that problem, this section presents several trace processing methods to separate unrelated events.

The goal of trace processing is to slicing an SoC execution trace into a set of sub-traces such that unrelated events are separated into different sub-traces as much as possible.  Effectiveness of trace slicing depends on available information from the traces.  In general, if more micro-architectural information is available for trace slicing, the mining performance would be better as more false dependencies among unrelated events can be eliminated.  Without loss of generality, two trace slicing techniques are described below. 

\paragraph{\bf Address Slicing}
This technique assumes that events carry memory addresses ${\tt addr}$ that are usually available for memory related tasks.  For example, the event of any downstream read or write flows contains the target address for read or write. 
Multiple flow instances may share the same target block address, therefore events with the same ${\tt addr}$ are not guaranteed to be correlated. However, when events carries different ${\tt addr}$ values, they are always unrelated, thus can be separated.
For example, consider the following trace over $e_1, e_2$ and $e_3$.
\begin{equation}
(\{e_1(10)\},  \{e_2(10),e_1(15)\} , \{e_3(10),e_2(15)\}, \{e_1(15)\})
\label{eq:trace-2}
\end{equation}
Each bracket indicates the set of events executed at the same clock cycle, and the value inside the parentheses indicates the value of the ${\tt addr}$ data filed. For example, $e_1(10)$ represents the instance of event $e_1$ with address $10$. 
Address slicing slices such trace into two sub-traces as shown below with respect to different ${\tt addr}$ values.  In the resulting sub-traces, all events in a single trace carry the same {\tt addr} value. 
\begin{align*}
addr=10: &\ \ \ \ \ \ \{\{e_1\},  \{e_2\} , \{e_3\}\} \\
addr=15: &\ \ \ \ \ \ \{ \{e_1\} , \{e_2\}, \{e_1\}\}
\label{trace-addr}
\end{align*}

\paragraph{\bf Causality Slicing}
This trace slicing technique considers the causality property discussed in \textit{Definition 6} in Section~\ref{sec:rules}, and tries to separate events that do not follow causality property into different sub-traces. 
This algorithm starts with an execution trace and an empty set $st$ to hold sub-traces. Then, it iterates through each individual event $e_x$ in the trace, and check if it satisfies the causality property with any existing sub-traces. 
Let $(e_0,\ldots, e_i)$ be a sub-trace.  If $e_i.dest = e_x.src$, $e_x$ is added to the back of the sub-trace. Otherwise,
a new sub-trace is added to the set $st$ with $e_x$ inserted.

To illustrate the basic idea of causality slicing, consider the simple example trace $\rho$ below
\[
\rho = (e_0, e_1, e_2, e_3).
\]
The source and destination of each event is shown in Table~\ref{table:cas-inf}. The destination of $e_0$ matches with the source of $e_2$. As the same way, $e_1$ matches with $e_3$, forming to two sub-traces, as shown below on the right.
\begin{table}[ht]
\begin{minipage}{.5\linewidth}
\centering{
\begin{tabular}{|c|c|c|} 
\hline
Event &Src &Dest \\
\hline
$e_0$ &A&B\\
\hline
$e_1$ &D&E\\
\hline
$e_2$ &B&C\\
\hline
$e_3$ &E&F\\
\hline
\end{tabular}
}
\vspace{3mm}
\caption{}
\label{table:cas-inf}
\end{minipage}%
\begin{minipage}{.5\linewidth}
\centering{
\begin{tabular}{ cc} 
$\rho_0$: & $(e_0, e_2)$\\
$\rho_1$: & $(e_1, e_3)$\\
&\\
&
\end{tabular}
}
\end{minipage}
\end{table}

This example shows the ideal scenario for the causality slicing. However, one event may satisfy the causality property with multiple sub-traces.
When this situation happens, and only limited information is available to determine which sub-trace an event belongs to, those sub-traces are combined into a single sub-trace, and that event is added to the combined sub-trace. This is to maintain the original temporal orderings, and at the same time to avoid introducing false dependencies among events.
\subsection{LSTM Training}

Neural networks have shown its significant impacts in various fields in recent years~\cite{hinton1990introduction}. 
In this paper, we explore the recurrent neural networks (RNNs), a special type of neural networks that is designed to capture sequential dependencies.


\smallskip
\noindent\textbf{Architecture. }
We use a particular type of RNNs, \emph{Long Short Term Memory (LSTM)} networks, in this paper. 
An LSTM is able to capture the ``long-term dependencies" compared to regular RNNs. The basic LSTM unit is composed of a cell, an input gate, an output gate, and a forget gate. 
At each time step, an LSTM block uses these gates to
decide the portions of the information to retain and update its state, and to produce the new output $h_t$ for the connected block. 
Each LSTM unit contains a set of weights that controls how the gates operate.  Training an LSTM model is the process of assigning the proper values to the weights of the network.  For this work we use
the categorical cross-entropy loss function to calculate the model errors which are then used to update the weights of each unit.

\begin{figure}
\centering{
\resizebox{.45\textwidth}{!}{
\includegraphics[]{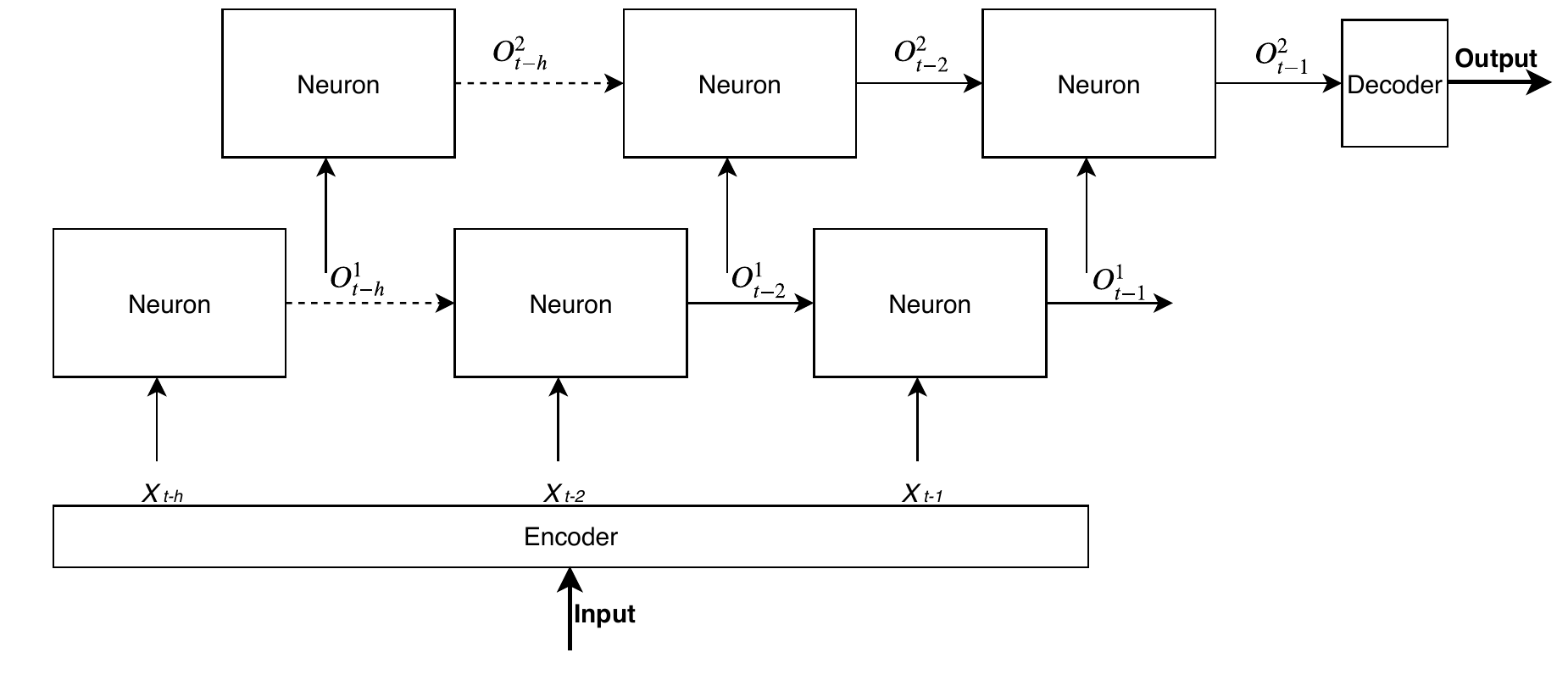}
}
\caption{The overall structure of the proposed LSTM model}
\label{fig:LSTM}
}
\end{figure}

Fig.~\ref{fig:LSTM} shows the structure of the LSTM models used in Fig.~\ref{fig-overview}. It is composed of two hidden layers,
an input layer and an output layer, using standard encoding-decoding algorithms. 
Every layer contains a set of recurrent LSTM units, one for each unique event in the traces for training.

The input layer of an LSTM model encodes each unique event in an input trace with an one-hot vector $\vec{x}$.
Then, as the hidden layer updates its states, the LSTM model calculates the likelihood of an event being executed in the next time step. 
The output layer decodes the final output into a probability distribution function using a standard multinomial logistic function. The probability distribution is represented as a set of $P(e_{h+1}\ |\ w)$ for each unique event $e_{h+1}$ where  $w$ is a sequence of events $(e_0, \ldots ,e_h)$.  


\smallskip
\noindent\textbf{Training.~}
In the proposed approach, the patterns are grouped by their lengths (the numbers of events in a pattern). We create a set of LSTM models, one for patterns of a fixed length.  
As the desired patterns are often unknown, this approach allows flexibility in various patterns that can be mined.  Pattern lengths for mining can be specified by the user. 
 
For a LSTM model targeting at $w-$length patterns, a set of pairs $(S, e_w)$ are generated from the input traces for training. $S$ is a sub-sequence of $w-1$ events occurred in $w-1$ time steps before $e_w$.  $S$ is used as an input, while $e_w$ is used as the output label for $S$.   
Let $\rho$ be a sequence of $k$ events as our training trace:
\[
\rho = (e_0, e_1 \ldots e_{k-1})
\]
This approach considers the set of all pairs of $(S,e_w)$ out of $\rho$. 
For this example, it includes

$\{(e_0 \ldots e_{w-2}), e_{w-1})$， 
\ldots $(e_{k-w} \ldots e_{k-2}\}, e_{k-1})\}$


\subsection{Pattern Extraction}
\label{sec:patext}

After all LSTM models are properly trained, a pattern extraction process is then applied to each model in sequence, as shown in Fig.~\ref{fig-overview}. For a model trained for pattern length $w$, it takes a sequence of $w-1$ events as input, and computes the output probability distribution for all unique events. 
We use $\theta$ to define the minimum probability for a sequence of events to be considered as a pattern. Consequently, our framework only extracts patterns whose probabilities exceed the predefined threshold $\theta$.

In order to extract all valid patterns, we need to consider all possible input event sequences $S$, and check their output probability distributions. However, as the pattern length increases, the number of such sequences for consideration grows exponentially, leading to high runtime complexity. 
To address this problem, we propose to chain the models in the ascending order with regard to their pattern lengths so that valid patterns extracted from one model are used as inputs of the next model for extracting longer patterns.
Only the first model considers all input sequences of single events.
In this setup, if a sequence is not identified as a pattern, then any extension of such sequence will not be considered as a pattern. 




\section{Experiments}
\label{section:results}



\subsection{Setup}

\begin{figure}
\begin{center}
\resizebox{2.5in}{!}{
\includegraphics[width=\linewidth]{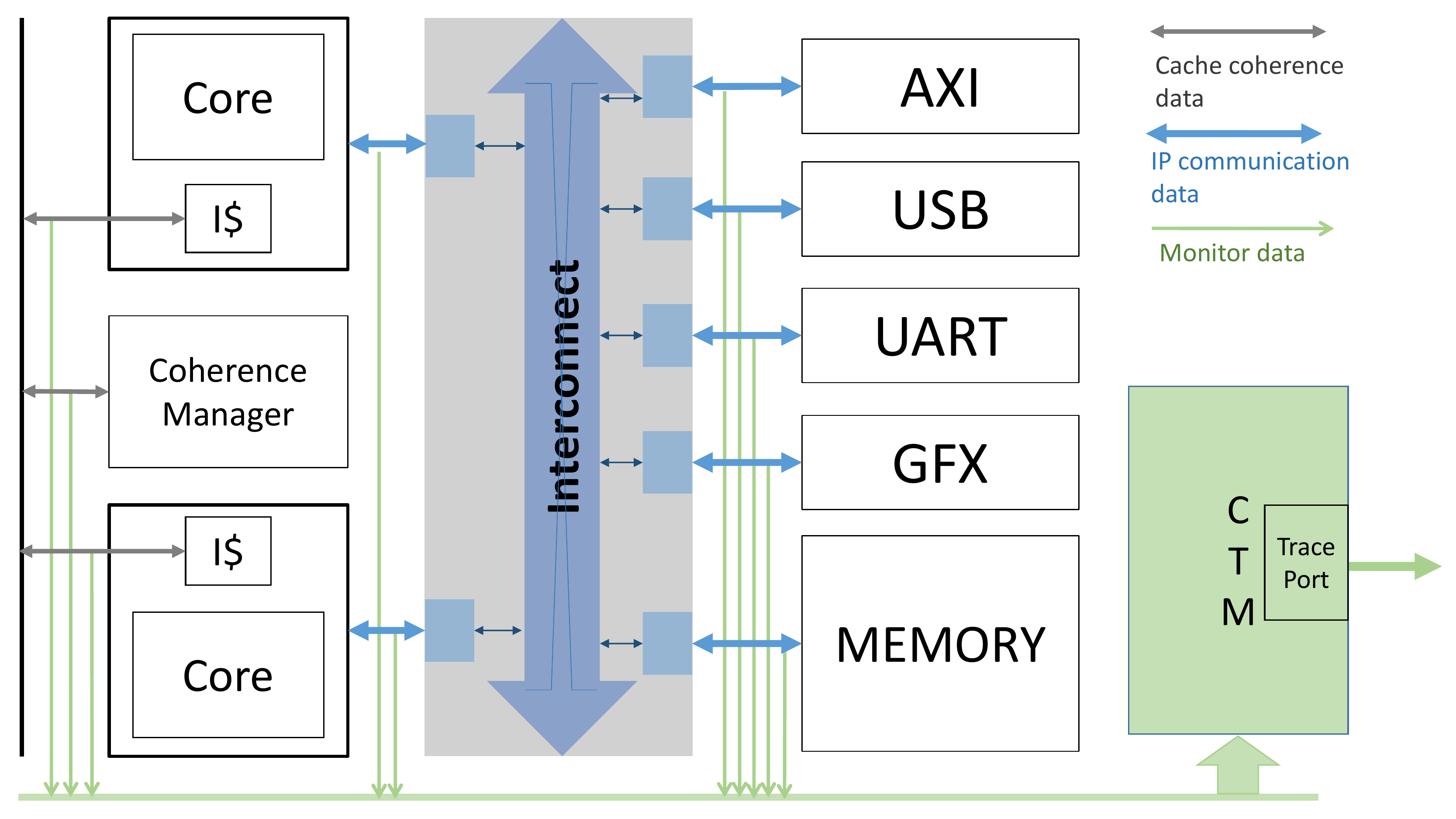}
}
\caption{An SoC prototype where each communication link is attached with a monitor. There can be multiple links between a pair of components.}
\label{rtlstruc}
\end{center}
\end{figure}

To evaluate the presented framework, a non-trivial SoC model is developed, as shown in Figure~\ref{rtlstruc}.
We implement a total of $10$ message flows in this model, including cache coherence, downstream read/write protocols for CPUs, upstream read/write for the peripheral blocks, \emph{etc}.  Those flows, although simplified, are adapted from real industry designs.
The model is simulated in a random test environment where CPUs and four other peripheral blocks are programmed to randomly select a flow to initiate with a delay between 1 to 10 cycles.  
In a simulation run, each of those blocks activates {100} flow instances, and a total of $500$ flow instances are activated during the entire simulation.  We demonstrate the proposed message flow mining framework on this SoC model by showing various patterns it can extract.  We also present the effects of several post-processing techniques that show great potential in filtering invalid patterns.

%
\subsection{Results}
\label{sec:exp-results}

Two hundred execution traces of the SoC model are collected for experiments.  Each trace is generated by simulating the SoC model with a different random seed so that a diverse set of traces are generated for training.  These execution traces are used to train the LSTM models with different pattern lengths. The training process is conducted on a GPU cluster with about 100 GPUs. The training process of each model takes approximately $20-30$ minutes. 

\begin{figure}
        
        \centering
        \resizebox{.5\textwidth}{!}{\includegraphics[width=\textwidth]{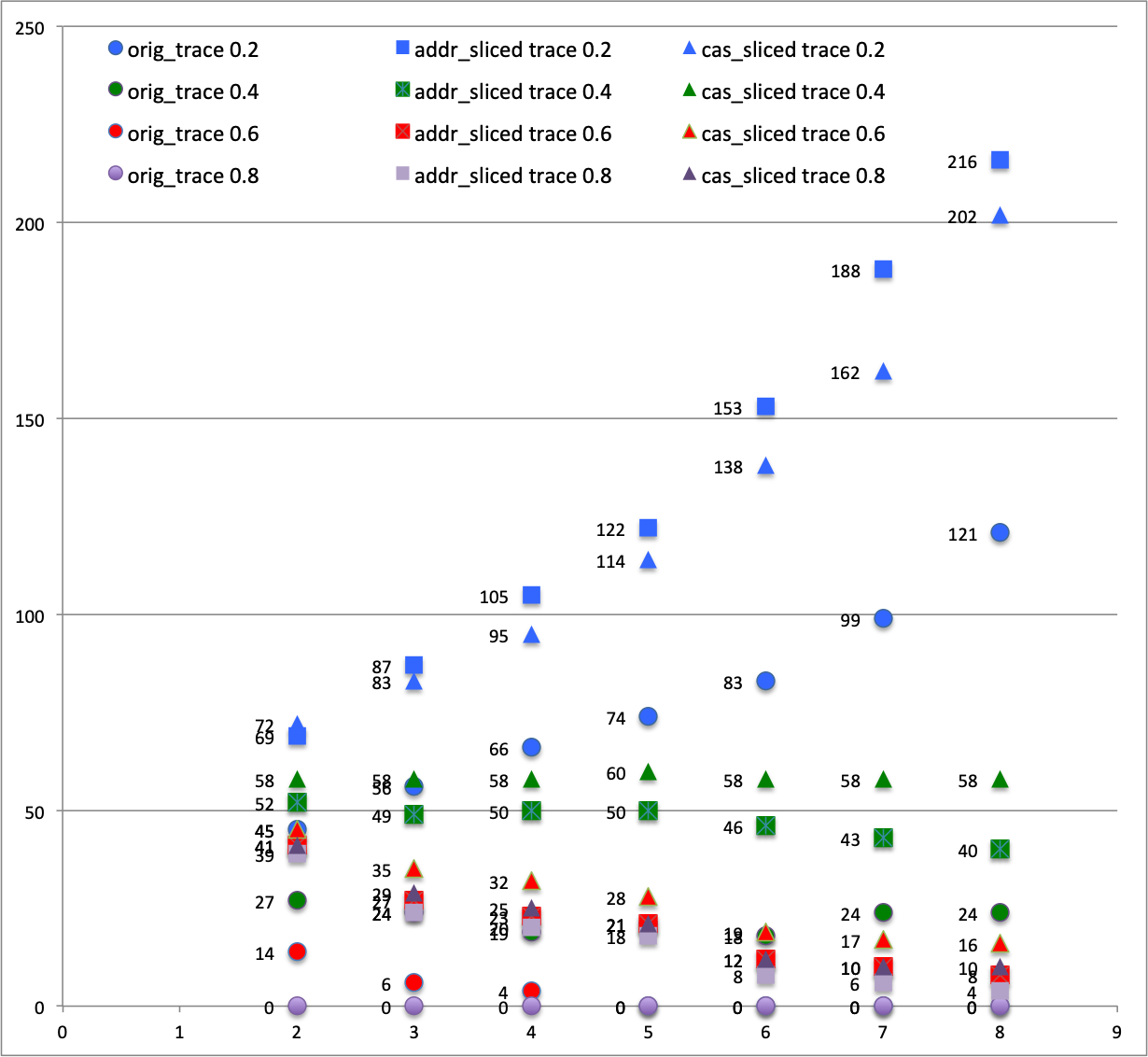}}
    \caption{Mined patterns of original traces and sliced traces with different thresholds $\theta$}
    \label{fig:all-results}
\end{figure}

Fig.~\ref{fig:all-results} shows the numbers of mined patterns from the original traces, address-sliced traces, and causality-sliced traces considering different probability thresholds for pattern extraction. The $x-$axis of Fig.~\ref{fig:all-results} represents the pattern lengths, and the $y-$axis represents the number of mined patterns for a pattern length.  The figure shows four sets of experimental results from using thresholds of $0.2$, $0.4$, $0.6$, and $0.8$, respectively.  The results from using a threshold are shown with an unique color.  The definition of $\theta$ can be found in Section~\ref{sec:patext}. 
It can be seen that the number of mined patterns significantly increases when a slicing technique is used. Moreover, when $\theta=0.8$, no patterns can be extracted from the original trace, while both sliced traces generate over 100 patterns with different lengths.
It also shows that less patterns are mined as the threshold increases. Every time the threshold increases by $0.2$, the mined patterns from the original traces reduce by more than $50\%$. Such reduction also occurs for the sliced traces, but at a much lower rate. 

Tables~\ref{table:patterns-orig} and \ref{table:patterns}
show the detailed results mined from the original traces, and sliced traces using both slicing techniques,  respectively, with the threshold $\theta=0.2$. 
We use $V$ to represent valid patterns, $IV$ for Invalid patterns, $F$ for mined (found) patterns, and $NF$ for not-found patterns.  Note that the above notations are defined with respect to the ground truth patterns implemented for the SoC model.  
The comparison between Tables~\ref{table:patterns-orig} and \ref{table:patterns} shows the significant effect of trace slicing techniques. The total number of mined valid patterns increases from $28$ to $257$.  A side effect is that the many more invalid patterns are also mined. 
This is because that the slicing techniques eliminate many false dependencies in the original traces, leading to a more concentrated probability distribution among events remained in the sliced traces. As a result, more patterns, including both valid and invalid ones, can be mined.
In Tables~\ref{table:patterns-orig} and \ref{table:patterns}, the numbers in boldface are for the increased valid patterns or decreased invalided patterns, and the numbers in red are for the increased invalid patterns.

The results in Table~\ref{table:patterns} reveal two issues. First, the number of mined invalid patterns is much higher compared to the mined valid patterns, especially for patterns with longer lengths.
Second, the number of valid patterns that is mined is low considering the overall valid patterns.  We propose  techniques below to address those issues.

\paragraph{\textbf{Causality filtering}}
 This technique leverages the causality property in Definition~\ref{def:causality} to avoid generating patterns that violate the property. Specifically, given an input sequence $(e_1, \ldots, e_{w-1})$, even if the trained LSTM model returns event $e_w$ such that $P(e_w|e_1, \ldots, e_{w-1})$ is above the given threshold, sequence $(e_1, \ldots, e_{w-1}, e_w)$ will not be returned as a pattern if $e_{w-1}$ and $e_w$ do not meet the causality property.  
 Table~\ref{table:patterns-cas} shows that this technique can lead to significant reduction in the numbers of mined invalid patterns compared with the results on row $IV\&F$ in Table~\ref{table:patterns}.

\paragraph{\textbf{Additional threshold for model input generation}}
Our approach uses a threshold to avoid invalid patterns during the pattern extraction. 
However, it fails to consider the case where a low probability sequence could be a prefix of a longer pattern with sufficiently high probability. 
To address this issue, 
we use an additional threshold $\theta'$ such that $\theta' < \theta$ to generate candidate input sequences for extracting longer patterns. 
Table~\ref{table:patterns-thre} shows the results from using this technique where $\theta' = 0.05$ and $\theta$ remains at $0.2$.  The number of mined valid patterns in Table~\ref{table:patterns-thre} increases to $408$ in total from $261$ in Table~\ref{table:patterns-cas}.
However, as more sequences are considered for pattern extraction in each step, more invalid patterns are also mined as shown by the numbers in red. 

\paragraph{\textbf{Initiating Event Filtering}}

According to \textit{Definition 2}, an execution of a message flow is a sequence of events from the initial state. As a result, a pattern that represents an execution of a flow should start with an initiating event. 
If the initiating events can be identified from execution traces, we can limit the scope of valid patterns to be mined. In the following, we show a simple method to identify the initiating events from the execution traces.  Given a set of traces $\Phi$ for pattern mining, event $e$ is an initiating event if $e.{\tt src} \not= e'.{\tt dest}$ for all $e' < e$ in $\rho$ for all $\rho \in \Phi$.  


Once all initiating events are identified using this technique, our approach can return patterns started with those initiating events. Among all ground truth patterns of the implemented message flows, this technique allows $2$ out of $4$ two-event patterns, $5$ out of $10$ four-event patterns, and $23$ out of $44$ eight-event patterns to be extracted.  

\begin{table}
{\footnotesize
\centering{
\begin{tabular}{ |c|c|c|c|c|c|c|c|} 
 \hline
 Length &2&3&4&5&6&7&8\\
  \hline
   V\&F&21&6&1&0&0&0&0\\
  \hline
IV\&F&24&50&65&74&83&99&121\\
  \hline
V\&NF&97&140&139&130&118&88&44\\

\hline
 \end{tabular}
 }
\vspace{2mm}
\caption{Mined patterns of the original trace with $\theta = 0.2$}
\label{table:patterns-orig}




\centering{
  \begin{tabular}{ |c|c|c|c|c|c|c|c|} 
 \hline
 Length &2&3&4&5&6&7&8\\
  \hline
   V\&F& ${\bf 56}$ &${\bf 45}$& ${\bf 36}$& ${\bf 34}$ & ${\bf 39}$ & ${\bf 32}$& ${\bf 19}$\\
  \hline
IV\&F& ${\bf 16}$ & ${\bf 38}$ & ${\bf 59}$ & ${\bf \color{red}{80}}$ & ${\bf \color{red}{99}}$ & ${\bf \color{red}{130}}$ & ${\bf\color{red}{183}}$\\
  \hline
V\&NF&62&101& 104 &96& 79 &56&25\\
\hline
 \end{tabular}
 }
\vspace{2mm}
\caption{Mined patterns of the sliced trace with $\theta = 0.2$}
\label{table:patterns}

\centering{
\begin{tabular}{ |c|c|c|c|c|c|c|c|} 
 \hline
 Length &2&3&4&5&6&7&8\\

  \hline
   V\&F&56&45&36&34&{39}&32&19\\
  \hline 
New IV\&F & {\bf 4} & {\bf 14} & {\bf 21} & {\bf 31} & {\bf 37} & {\bf 44} & {\bf 66}\\

  \hline
V\&NF&62&101&104&96&79&56&25\\
\hline
 \end{tabular}
 }
 \vspace{2mm}
\caption{Mined patterns after applying causality filtering}
\label{table:patterns-cas}

\centering{
 \begin{tabular}{ |c|c|c|c|c|c|c|c|} 
 \hline
 Length &2&3&4&5&6&7&8\\
  \hline
  V\&F&{\bf 79}&{\bf 80}&{\bf 70}&{\bf 57}&{\bf 57}&{\bf 42}&{\bf 23}\\
  \hline
IV\&F&{\bf\color{red} 8}&{\bf\color{red} 31}&{\bf\color{red} 59}&{\bf\color{red} 89}&{\bf\color{red} 108}&{\bf\color{red} 120}&{\bf\color{red} 146}\\
  \hline
V\&NF&{\bf 39}&{\bf 66}&{\bf 50}&{\bf 73}&{\bf 61}&{\bf 46}&{\bf 21}\\
\hline
 \end{tabular}
 }
 \vspace{2mm}
\caption{Mined patterns after applying additional threshold}
\label{table:patterns-thre}



}
\end{table}

\section{Conclusion}
\label{sec:conclude}

This paper presents an approach that automatically extracts message flow specifications from SoC transaction-level traces.  It includes several trace processing techniques that reduce false sequential dependencies while preserving the essential ones in SoC execution traces.  It utilizes the innovative LSTM models to capture sequential dependencies in the traces for pattern extraction. 
We apply this approach on the execution traces of a non-trivial multicore SoC model, and evaluate quality of mined specifications.  
In the future, we plan to optimize it to consider various dynamic information available in a event, and further improve the quality of extracted specifications by reducing invalid patterns while increasing valid patterns.

\section*{Acknowledgment}

The work presented in this paper is partially supported by a gift from the Intel Corporation, and grant \textit{{3910-1009-03}} from CyberFlorida.
The authors would like to thank anonymous reviewers for their constructive comments.


\end{document}